# MCD64A1 Burnt Area Dataset Assessment using Sentinel-2 and Landsat-8 on Google Earth Engine: A Case Study in Rompin, Pahang in Malaysia

Yee Jian Chew, Shih Yin Ooi, Ying Han Pang
*Faculty of Information Science and Technology*
*Multimedia University*
Melaka Malaysia
chewyeejian@gmail.com, syooi@mmu.edu.my, yhpang@mmu.edu.my

*Abstract*— This research paper intends to explore the suitability of adopting the MCD64A1 product to detect burnt areas using Google Earth Engine (GEE) in Peninsular Malaysia. The primary aim of this study is to find out if the MCD64A1 is adequate to identify the small-scale fire in Peninsular Malaysia. To evaluate the MCD64A1, a fire that was instigated in Rompin, a district of Pahang on March 2021 has been chosen as the case study in this work. Although several other burnt area datasets had also been made available in GEE, only MCD64A1 is selected due to its temporal availability. In the absence of validation information associated with the fire from the Malaysian government, public news sources are utilized to retrieve details related to the fire in Rompin. Additionally, the MCD64A1 is also validated with the burnt area observed from the true color imagery produced from the surface reflectance of Sentinel-2 and Landsat-8. From the burnt area assessment, we scrutinize that the MCD64A1 product is practical to be exploited to discover the historical fire in Peninsular Malaysia. However, additional case studies involving other locations in Peninsular Malaysia are advocated to be carried out to substantiate the claims discussed in this work.

## I. INTRODUCTION

With the increasing availability of public satellite imagery [1], the remote sensing field has been revolutionized and opened up many new research directions for environmental monitoring, disaster management, and urban planning. Referring to Vatsavai et al. [2], they mentioned that Google produced approximately 25 petabytes (PB) every day in 2012, where a large proportion of the data are spatiotemporal data. Following the projection of an increase of 20% in geospatial data annually [3], it is postulated that traditional approaches to processing and analyzing the data will be very challenging [4]. Hence, several big data platforms such as Google Earth Engine (GEE) [5] and Planetary Computer [6] have been developed by Google and Microsoft to access, analyze, and visualize remote sensing data in the cloud environment. By easing most of the processing resources from the local environment (i.e., not required to download, process, and analyze the tremendous volume of satellite data locally), researchers can devote their work to focusing on analyzing and developing solutions for their research problems without vast concern on the availability of local computation and storage resources [7].

In this paper, GEE is utilized to assess the burnt area detected by the MCD64A1 burnt area product in Peninsular Malaysia. Although several studies had been carried out to assess and validate the burnt area detected by MCD64A1, most of the works either committed to assessing the burnt area at a global scale [8] or at a different geographical region [9]–[15]. Aside from the distinct geographical locations, the majority of the past research focused only on large-scale fires. Though there is no specific definition for a large fire, government agencies and politicians from the US consider any fire that burnt an area greater than 10,000 ha as a large-scale wildfire [16]. Therefore, it is vital to emphasize that most of the fires that occurred in Peninsular Malaysia were considered small-scale (< 100 ha) [17]. Thus, it is essential to assess whether the MCD64A1 is suitable to be employed to pinpoint the burnt area in Peninsular Malaysia. To the best of our knowledge, this is the first initiative that adopted GEE to evaluate the MCD64A1 burnt area in Peninsular Malaysia.

The fire that was triggered in Rompin, Pahang on March 2021 [18]–[20] had been chosen as our case study in this work. Details of the region of interest and fire episode are deliberated in Section II. As the Malaysian government does not publish the affected fire perimeters, it is impossible to compare the burnt area detected by MCD64A1 burnt area with the field study. Hence, we refer to the public news source [18]–[20] to obtain further details associated with the fire occurrence. Additionally, the burnt area is also assessed by utilizing the remote sensing imagery from Sentinel-2 and Landsat-8. Information related to the methodology and the discussions of experimental results are delivered in Section III and Section IV.

## II. STUDY AREA

Rompin, a district in the state of Pahang, Malaysia is selected as the study area. Based on the analysis of Fire Information for Resource Management System (FIRMS) hotspots from 2001 to 2021 in Pahang, Rompin was identified as a high-risk forest fire region due to the great number of hotspots detected [21]. Several fire incidents had been reported in March 2021 in Rompin [18]–[20]. Though a few smaller fire incidents were reported in the study area, the investigation in this paper is predominantly focused on assessing the burnt area in Muadzam Shah, Rompin. This is because the burnt area reported by the news [18]–[20] was larger than 300 ha. The region of interest is depicted in Fig. 1, and the approximated centered coordinates point is located at 2.9469°N, 103.2774°E.

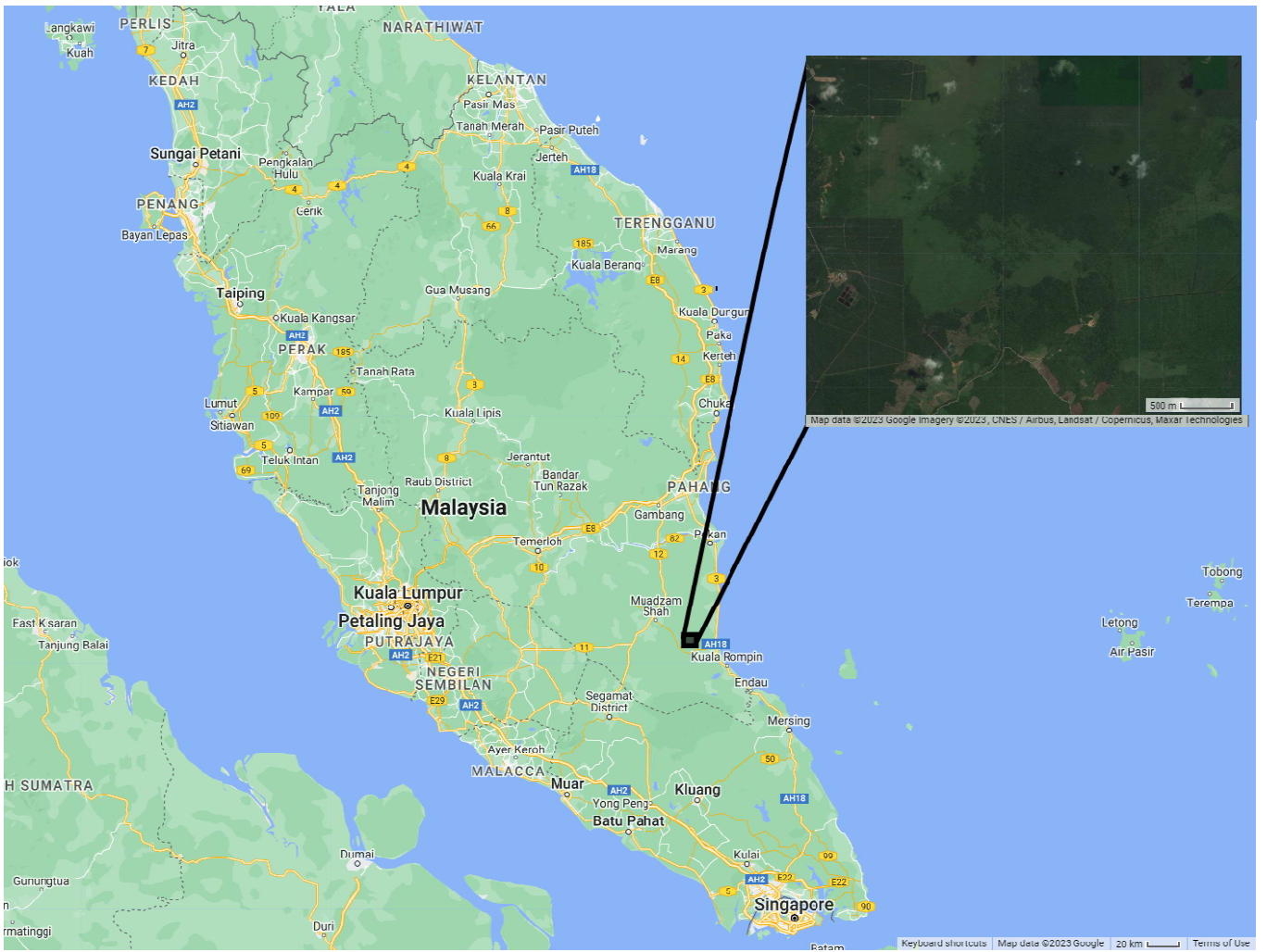

Fig. 1. Region of Interest (Study Area)

## III. METHODOLOGY AND DATASETS SELECTION

### A. Datasets

Since we are utilizing the GEE platform, the discussions will be concentrated on the available burnt area datasets in GEE, specifically FireCCI51 [22], Globfire Fire Event [23], and MCD64A1 [24]. The resolutions, temporal, and spatial coverage for the three datasets are listed in Table 1. As the fire in Rompin was sparked in March 2021, only MCD64A1 burnt area product is suitable to be selected for our work due to its temporal availability.

TABLE I. BURNT AREA DATASETS DESCRIPTION

| Dataset | Temporal Coverage | Resolutions | Spatial Coverage |
|---|---|---|---|
| FireCCI51 v5.1 | 2001-01-01 to 2020-12-01 | 250 m | Global |
| Globfire Fire Event | 2001-01-01 to 2021-01-01 | Shapefiles | Global |
| MCD64A1.061 | 2000-11-01 to 2022-12-01 | 500 m | Global |

MCD64A1 [24] exploited the 500-m Moderate Resolution Spectroradiometer (MODIS) Surface Reflectance imagery to predict the date of fire incidents at 500-m grid cells for each MODIS active fire hotspot (i.e., FIRMS). In addition to the burnt area product, the FIRMS [25] hotspots, a near real-time thermal anomalies product that can detect fire location at 1 km is also adopted in this paper.

### B. Methodology

The burnt area detected by the MODIS64A1 product will be compared with the surface reflectance from Sentinel-2 and Landsat-8. The GEE script used to produce the true color images and burnt severity map of Sentinel-2 and Landsat-8 in Section IV are adapted from [26]. In general, the burnt severity map is generated by measuring the differenced Normalized Burn Ratio (NBR) by subtracting the post-fire NBR from the pre-fire NBR [26], [27]. NBR utilizes the shortwave-infrared (SWIR) and near-infrared (NIR) bands to project the burned area and severity. The pre-fire and post-fire images for the study area are produced by obtaining the images before and after the fire incidents from Sentinel-2 and Landsat-8, all the images for each of the satellites are then processed with the mosaic and clipped operation to create a seamless image for the region of interest for analysis. Since the primary contribution of our work centered on assessing the burnt area, we would like to bring the readers' attention to the details of the entire workflows and algorithms from [26].

TABLE II. INFORMATION AND PARAMETERS SETTINGS FOR SENTINEL-2 AND LANDSAT-8

| Dataset | Frequency | Temporal Coverage | Resolutions | Cloud Pixel (%) | Pre-fire Date | Post-fire Date | Total Images (Pre-fire) | Total Images (Post-fire) |
|---|---|---|---|---|---|---|---|---|
| Sentinel-2 Level-2A | 5 days | 2017-03-28 to 2023-03-01 | 10 m | Less than 25 % | 2020-02-01 to 2021-02-01 | 2021-03-01 to 2022-03-01 | 11 | 19 |
| Landsat-8 Collection 1 Tier 1 8-Day TOA Reflectance Composite | 16 days | 2013-04-07 to 2022-01-01 | 30 m | - | | | 22 | 17 |

Initially, the original script is run by just setting the timeframe of pre-fire and post-fire for the region of interest (Fig. 1) to produce the burnt severity map from the contiguous imagery obtained from Sentinel-2 and Landsat-8. The severity map produced by Landsat-8 imagery materializes greatly, however, the map produced by Sentinel-2 imagery turns out to be defective. Thus, the Image Viewer tool developed by [12], [28] is employed to access and view the Sentinel-2 imagery for our region of interest. From our observations, we notice that most of the images captured were very cloudy, and several of the images were corrupted. Hence, it can be deduced that the usability of the images might be discounted due to corrupted or cloudy images.

To resolve this issue, the original script is slightly modified to spawn a functional and usable image from Sentinel-2. First, we filtered all the images containing less than 25 % of cloudy pixels. Instead of using the Sentinel-2 Level-1C product, the Sentinel-2 Level-2A [29] product that had corrected the atmospheric effects is exploited. This is mainly due to the corrected product can produce a more accurate reflectance product that can manifest the burnt area more effectively. Our modified script can be directly accessed from the following Github repository *https://github.com/chewyeejian/GEE_BurntArea_MCD64A1_Pahang/*. It should be noted that users are required to import the script in GEE Code Editor to run the script.

IV. EXPERIMENTAL RESULTS AND ASSESSMENT

The true color imagery and burnt severity map are presented in Fig. 2 and Fig. 3, where Fig 2. utilizes Sentinel-2 imagery while Fig. 3 employed Landsat-8 imagery. Information and parameters for both Sentinel-2 and Landsat-8 that had been adopted in our work are tabulated in Table II.

Referring to Fig. 2(a) and Fig. 2(b), the pre-fire and post-fire mosaiced Sentinel-2 imagery after cloud masking was not exemplary due to the existence of thick clouds across most of the image. Despite that, it is sufficient to reveal that a fire had occurred in Rompin by inspecting the color contrast between the pre-fire and post-fire images. Similar observation can also be perceived from the mosaiced imagery produced by Landsat-8 in Fig. 3(a) and Fig. 3(b).

To assess the MCD64A1 burnt area product, the post-fire imagery is overlayed with the burnt area detected by MCD64A1 from 2021-03-01 to 2021-03-31. This range of dates is picked because of the fire that happened in March 2021 in the study area as described in Section II. From Fig. 2(c) and Fig. 3(c), it is apparent that the burnt area detected by MCD64A1 exhibits a strong correlation with the burnt area witnessed from the Sentinel-2 and Landsat-8 post-fire images. On the other hand, we also overlay the FIRMS hotspots detected in March 2021 with the post-fire imagery in Fig. 2(d) and Fig. 3(d). As the resolutions of the FIRMS hotspots dataset is lower than MCD64A1, the hotspots covered a larger area compared to the MCD64A1 burnt area product.

By leveraging the algorithm in [26], the burnt severity maps are generated in Fig. 2(e) and Fig. 3(e). Most of the moderate-low and moderate-high severity burnt areas correspond to the burnt areas revealed from the post-fire imagery and the MCD64A1 burnt area product. However, it should be noted that the burnt severity map produced by Sentinel-2 in Fig. 2(e) classified the clouds detected as moderate-low and moderate-high severity burnt areas. This is due to the limitation of the algorithm that adopted a change detection procedure to determine the burnt area [26]. From the results illustrated in Fig. 2 and Fig. 3, the MCD64A1 burnt area product significantly matched the burnt area observed from Sentinel-2 and Landsat-8 imagery.

V. CONCLUSION

The burnt area assessment through Sentinel-2 and Landsat-8 imagery using GEE provided valuable insights into the usability of the MCD64A1 burnt area product for the study area. Since the MCD64A1 product displays a strong correlation with the burnt area detected from Sentinel-2 and Landsat-8, we scrutinize that the MCD64A1 is feasible to identify the historical fire incidents instigated in Peninsular Malaysia. In the future, a follow-up analysis can be performed to analyze the factors of forest fire by incorporating the meteorological variables, topology variables, etc. However, additional case studies are recommended to be conducted to supplement the practicality of the MCD64A1 product to detect small-scale forest fires in Malaysia.

ACKNOWLEDGMENT

This research work was supported by a Fundamental Research Grant Schemes (FRGS) under the Ministry of Education and Multimedia University, Malaysia (Project ID: FRGS/1/2020/ICT02/MMU/02/2), and Chey Institute for Advanced Studies (ISEF).

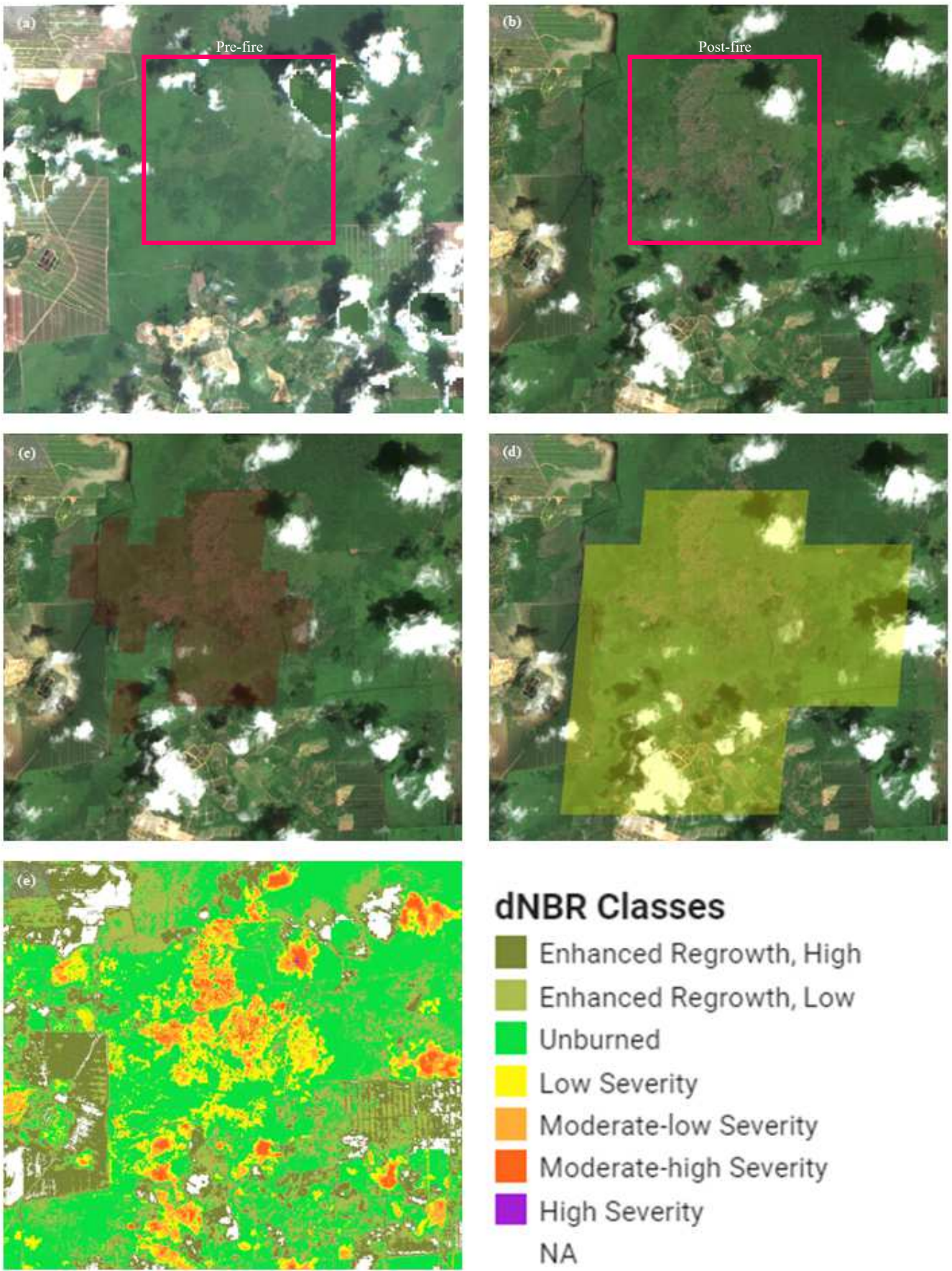

Fig. 2. Wildfire in Rompin, Pahang to illustrate the results from Sentinel-2. (a) Pre-fire mosaiced cloud-mask Sentinel-2 image, (b) Post-fire mosaiced cloud-mask Sentinel-2 image, (c) Post-fire mosaiced cloud-mask Sentinel-2 image with MCD64A1 Burnt Area product, (d) Post-fire mosaiced cloud-mask Sentinel-2 image with FIRMS hotspots, (e) dNBR

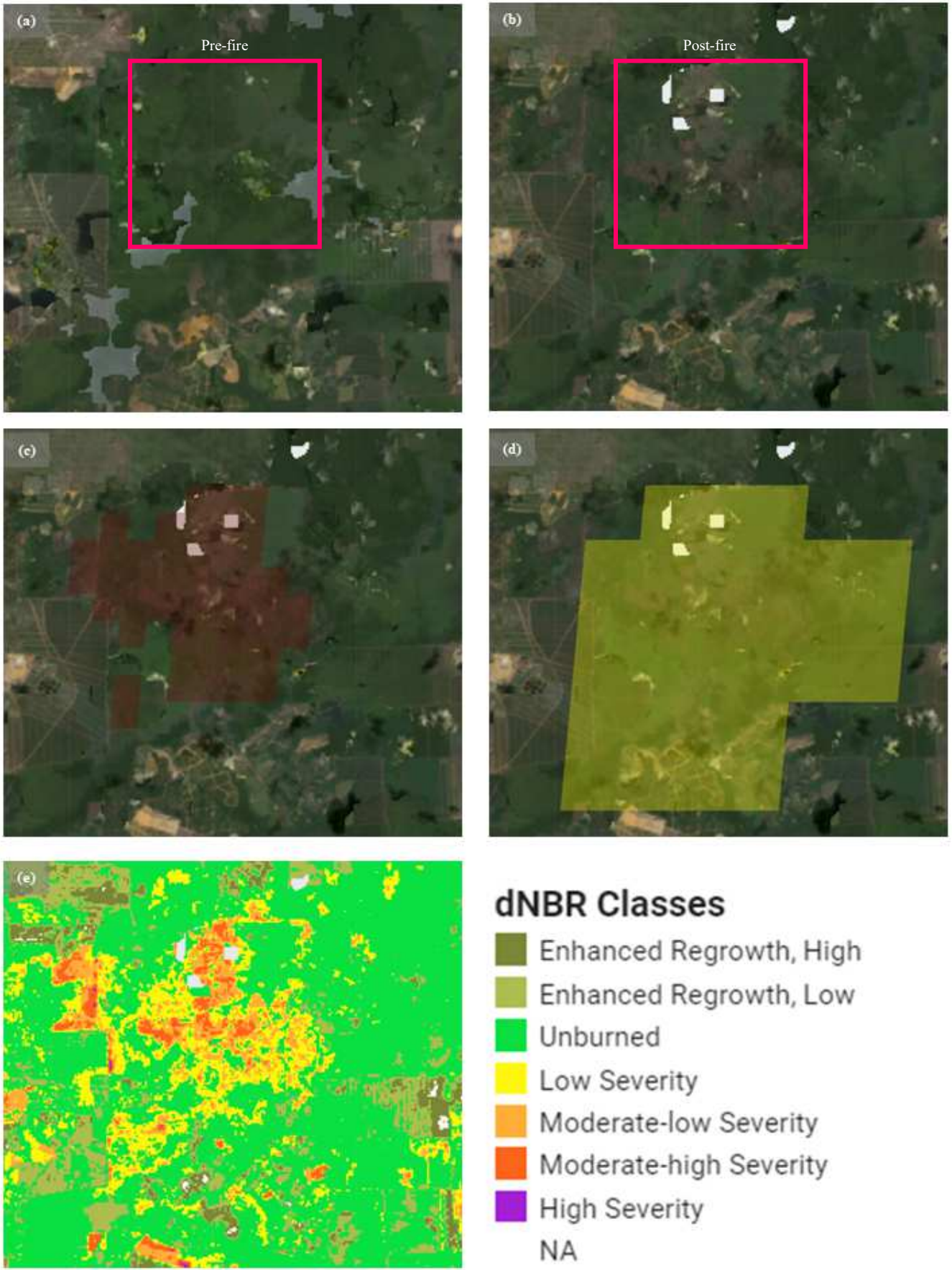

Fig. 3. Wildfire in Rompin, Pahang to illustrate the results from Landsat-8. (a) Pre-fire mosaiced cloud-mask Landsat-8 image, (b) Post-fire mosaiced cloud-mask Landsat-8 image, (c) Post-fire mosaiced cloud-mask Landsat-8 image with MCD64A1 Burnt Area product, (d) Post-fire mosaiced cloud-mask Landsat-8 image with FIRMS hotspots, (e) dNBR